\documentclass[twocolumn,aps,floatfix,superscriptaddress]{revtex4-1}
\usepackage{amsmath,amssymb,eucal,graphicx,float,epstopdf,fancyhdr}

\usepackage{ulem}
\usepackage{color}

\begin{document}

\title{Signal optimization in urban transport:\\
A totally asymmetric simple exclusion process with traffic lights}
\def\shorttitle{Signal optimization in urban transport} 

\author{Chikashi Arita}
\affiliation{Theoretical Physics, Saarland University, 66041 Saarbr\"ucken, Germany}
\author{M. Ebrahim Foulaadvand} 
\affiliation{Department of Physics, University of Zanjan, P. O. box 45196-313, Zanjan, Iran}
\affiliation{School of Nanoscience, Institute for Research in Fundamental Sciences (IPM), P. O. Box 19395-5531, Tehran, Iran}
\author{Ludger Santen}
\affiliation{Theoretical Physics, Saarland University, 66041 Saarbr\"ucken, Germany}
\def\shortauthor{C. Arita, M. E. Foulaadvand, L. Santen}

\pagestyle{fancy}
 \lhead{\shorttitle} \chead{} \rhead{\shortauthor} 
 \renewcommand{\headrulewidth}{0.5pt} 
 \renewcommand{\footrulewidth}{0.5pt} 
 \lfoot{} \cfoot{\thepage} \rfoot{} 
 
 \fancypagestyle{titlepage}{
 \renewcommand{\headrulewidth}{0pt} 
 \renewcommand{\footrulewidth}{0.5pt} 
 \lhead{} \chead{} \rhead{}
 \lfoot{} \cfoot{\thepage} \rfoot{} }

\begin{abstract}
 We consider the exclusion process on a ring with time-dependent defective bonds at which the hopping rate periodically switches between zero and one. This system models main roads in city traffics, intersecting with perpendicular streets. We explore basic properties of the system, in particular dependence of the vehicular flow on the parameters of signalization as well as the system size and the car density. We investigate various types of the spatial distribution of the vehicular density, and show existence of a shock profile. We also measure waiting time behind traffic lights, and examine its relationship with the traffic flow.
\end{abstract}

 \maketitle
\thispagestyle{titlepage}

\section{Introduction\label{sec:intro}} 
We often experience traffic jams during rush hours in cities. In urban networks, traffic flows are controlled by traffic lights. Ideally, the cycles of the traffic lights should be coordinated in a way that optimizes the travel times in the network or avoids deadlock situations. The motivation of this work is to explore systematically the optimization of traffic flow, by using a simple transport model. In traffic engineering excluded-volume effect and stochastic fluctuations are usually not taken into account. The totally asymmetric simple exclusion process (TASEP) \cite{bib:S:70,bib:Derrida:98} is a minimal model that includes these features. The TASEP is one of cellular-automaton models with stochastic time evolution, which are systems of interacting particles on lattices. In the TASEP, each site of the lattice is either occupied by a particle or empty, and each particle stochastically hops to the right neighboring site, if this target site is empty. Undoubtedly the TASEP has played a prominent role as a paradigmatic model for describing many driven non-equilibrium systems, especially physics of transport phenomena \cite{bib:Derrida:98,bib:SCN:11}. Since its introduction for theoretical description of the kinetics of protein synthesis \cite{bib:MGP:68}, the TASEP has been generalized in many ways for e.g. describing biological transports, in particular the motion of molecular motors \cite{bib:PFF:03,bib:KL:04,bib:DSZ:07}. One of the intriguing disciplines which owe much to the TASEP is vehicular traffic flow \cite{bib:CSS:00}. Various features of traffic flow have been investigated in the framework of the TASEP such as overtaking \cite{bib:Foolad:00}, intersection of streets \cite{bib:FN:07}, flow at junctions \cite{bib:HJHW:08}, queuing process \cite{bib:Arita:09}, anticipation effect \cite{bib:HJHW:10}, time and spatial headway at intersections \cite{bib:KH:11}, on-ramp simulation \cite{bib:XLS:12}, pedestrian-vehicles flow \cite{bib:HA-R:12,bib:IN:14}, roundabout \cite{bib:FM:16} and shortcut \cite{bib:BPB:15}. Models of traffic flow at intersections have been also investigated by other approaches than the TASEP, mainly in discrete-time frameworks \cite{bib:FSS:04,bib:HWL:11,bib:VCTMRZV:09,bib:BF:09,bib:FFB:10,bib:JWJ:16,bib:ELE-ZB:16}.

In order to investigate the effects of traffic lights, one can introduce time-dependent hopping probabilities in some particular sites of lattice traffic models. For example in \cite{bib:BBSS:01}, discrete-time models were analyzed on regular square lattices and some traffic-light strategies applied to optimize the flow in the system. On a simpler geometry i.e. ring, discrete-time models were also employed \cite{bib:HH:03,bib:Nagatani:06,bib:Nagatani:09} and fundamental diagrams (the curve of flow \textit{vs} the density of cars) were found to become constant when the density is in a certain range.

\begin{figure} 
\begin{center}
 \includegraphics[width=76mm]{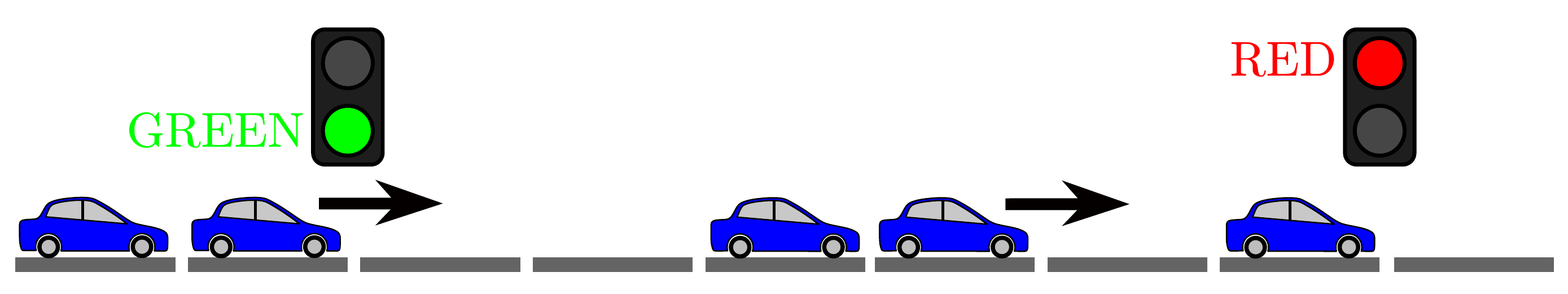}
\end{center}
\caption{\label{fig:illust}
Illustration of our model.
In the one dimensional lattice, each car move to the next site if it is empty with rate 1.
If a car is on a site just before a traffic light in red phase, the movement is not allowed. }
\end{figure}

In this work we focus on the control of traffic flows on a single main road of city networks, and analyze different strategies to optimize unidirectional flow by signalization. Specifically we use the continuous-time TASEP on a ring rather than more sophisticated discrete-time models \cite{bib:CSS:00}, which have been originally introduced for modeling highway traffic. Compared to these traffic models, in the continuous-time TASEP the cars' velocities fluctuate stronger. In our system, there is a traffic light which controls the conflicting flow of vehicles at each point intersected by a perpendicular street, see Fig.~\ref{fig:illust}. The traffic lights are regarded as local defects. As a special case, our model includes one of well-known inhomogeneous TASEPs, which was introduced by Janowsky and Lebowitz \cite{bib:JL:92}. We also remark that similar variants of the TASEP were introduced, e.g. \cite{bib:Wood:09} where the TASEP with time-dependent exit rate has been investigated and \cite{bib:TPPRC:13} where one site in the lattice can be blocked or unblocked stochastically for description of conformation changes in filamentous substrate. 

This work is organized as follows. We first (Sec.~\ref{sec:one}) analyze the case where there is only one traffic light on the street. We explore extensively its basic properties, mainly the fundamental diagrams, in various parameter regimes. We also consider various types of density profiles, according to averaging procedure. In particular the density profile by sample average \textit{converges} to a periodic function in time. (In the Appendix, we give a proof of the periodicity of physical quantities.) On the other hand, we observe a shock in the density profile by averaging over a large time window. Next (Sec.~\ref{sec:many}) we investigate the case where there are more than one traffic light. For simplicity traffic lights are equidistantly located. We treat two strategies for defining the difference between the offsets of two adjacent traffic lights: fixed and random ones \cite{bib:BBSS:01}. We examine the current, which depends on the strategies. We also measure the total waiting time of cars behind traffic lights, and explore its average and distribution in the two strategies. It turns out that there is an interrelation between the total waiting time and the current, in the case where the average distance a car drives in a period of lights and the distance between two adjacent lights are comparable. Finally (Sec.~\ref{sec:conclusions}) we summarize this work and mention possible future studies.

\section{TASEP with a single traffic light\label{sec:one}}
Let us consider first the TASEP with only one traffic light on a ring with $L$ sites. Each site is either empty or occupied by at most one particle (car). We denote the global density (i.e. the ratio between the number of cars and $L$) by $\rho$, and the occupation number of site $j$ at time $t$ by $\tau_j(t)$. The `hopping' rate of cars from site $j$ to $j+1$ is set to be 1. 
 Without loss of generality, we put the light between sites $L$ and 1. We assume that the light periodically changes its status from green to red and from red to green. We denote the cycle length (period) and the green phase ratio by $T$ and $ 0< g < 1 $, respectively, which are basic parameters in our model. The signal is green for $ gT$ unit of time and red for the rest of the cycle, i.e., $ (1-g)T$. 
More precisely, the jump rate from site $ L $ to 1 is given by the following time-dependent function $ \sigma (t) $: 
\begin{align}
\sigma (t) = \begin{cases}
 1 & (k \le \frac t T < k+g) , \\ 
 0 & (k+g \le \frac t T < k+1) , 
\end{cases}
\end{align}
where the integer $k = \lfloor t/T \rfloor $ is the cycle number. During the green phase, cars are allowed to cross the intersection, i.e. $ \sigma (t) = 1 $. If the signal is red, cars must wait behind the traffic light, i.e. $ \sigma (t) = 0 $, until the signal becomes green. Apparently the limiting case $g\to 1$ corresponds to the usual (homogeneous) TASEP on a ring, and $g\to 0$ to reflecting boundary conditions. In this work we consider only the case where $ g $ is not close to 0 or 1. 

One can define several types of density profiles.
The simplest one is the time average of the occupation number 
\begin{align}\label{eq:rhoj=}
 \rho_j = \frac{1}{t_2-t_1}\int_{t_1}^{t_2} \tau_j (t) dt ,
\end{align}
which is independent of time in the limit $t_2\to\infty$.
On the other hand, the average over independent samples (simulation runs) depends on time:
\begin{align}\label{eq:rhojt=}
 \rho_j (t) = \langle \tau_j (t) \rangle_{\rm sample} .
\end{align}
This converges to a \textit{periodically stationary} density profile
 $\rho^\text{p}_j (s) $ [satisfying $ \rho^\text{p}_j (T+s)= \rho^\text{p}_j (s) $] in the sense that 
\begin{align}
\label{eq:rhopjs}
 \lim_{ \kappa \to \infty } \rho_j ( \kappa T + s ) = \rho^\text{p}_j (s) 
\end{align}
 with $\kappa\in\mathbb Z_{\ge 0 }$. 
This is alternatively obtained by averaging over a time sequence in a single simulation run as
\begin{align}\label{eq:lim1/Ktauj(kT+s)}
 \rho^\text{p}_j(s) = \lim_{\kappa_2\to\infty} \frac{1}{ \kappa_2-\kappa_1}
 \sum_{ \kappa = \kappa_1}^{ \kappa_2-1} \tau_j ( \kappa T+s) 
\end{align}
 with $\kappa_1, \kappa_2 \in\mathbb Z_{\ge 0 }$. 
As a general remark, any time-dependent quantity $Q(t)$ as well as the density profile converges to a periodic function with the same period $T$, $Q^\text{p}(s) $, i.e. we have $ Q( \kappa T + s ) \to Q^\text{p}( s )$
 as $ \kappa \to \infty$ ($\kappa\in\mathbb Z_{\ge 0 }$). Actually the periodicity is predicted by Floquet's theory \cite{bib:Floquet:1883}. For convenience we give the proof in the Appendix. 

\begin{figure} 
\begin{center}
 \includegraphics[width=65mm]{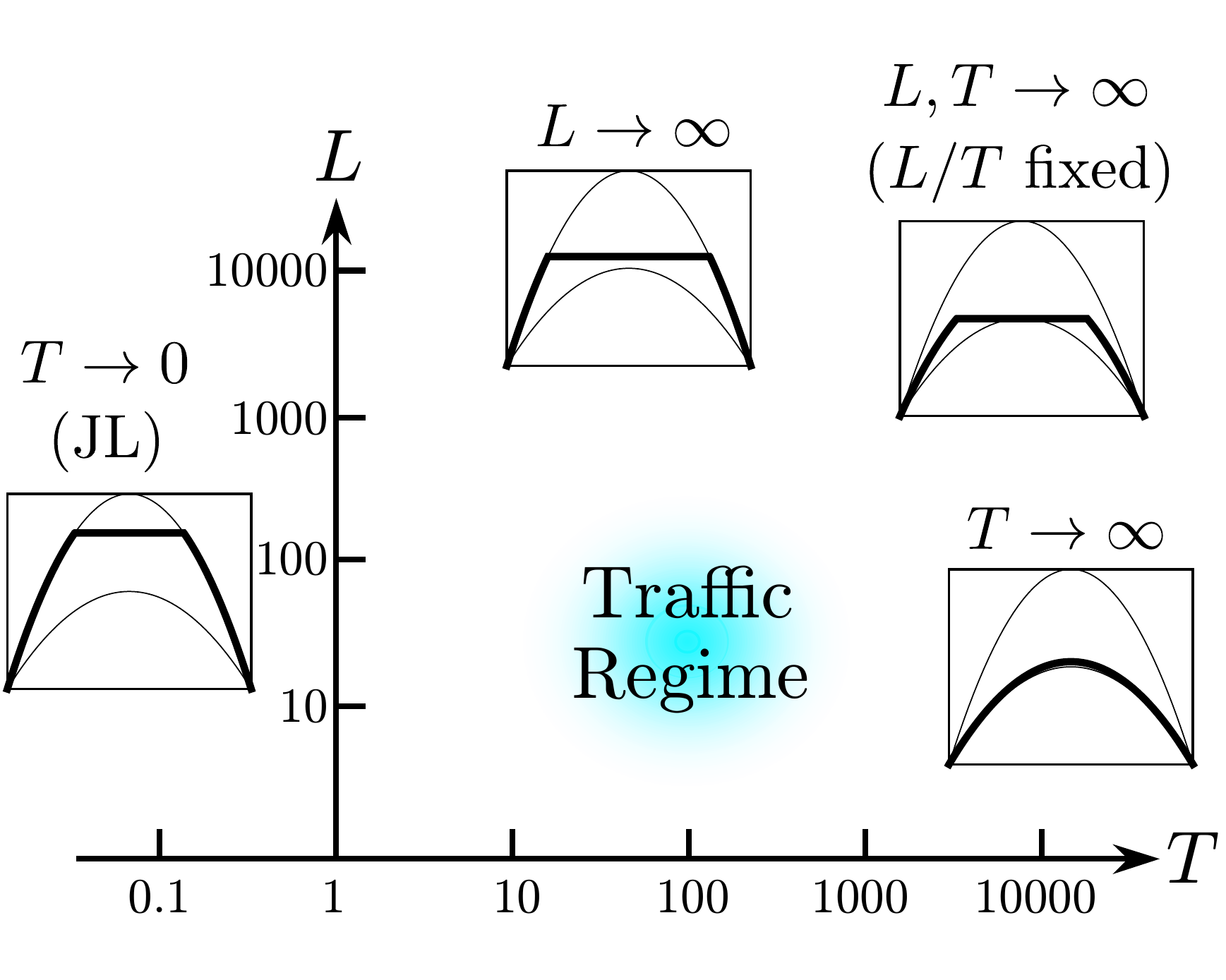}
\caption{\label{fig:limits} Limits of the fundamental diagram.
The thin lines are parabolas $J=\rho (1-\rho ) $ and $J=g\rho (1-\rho ) $.}
\end{center}
\end{figure}

\begin{figure}
 \begin{center}
 \includegraphics[width=76mm]{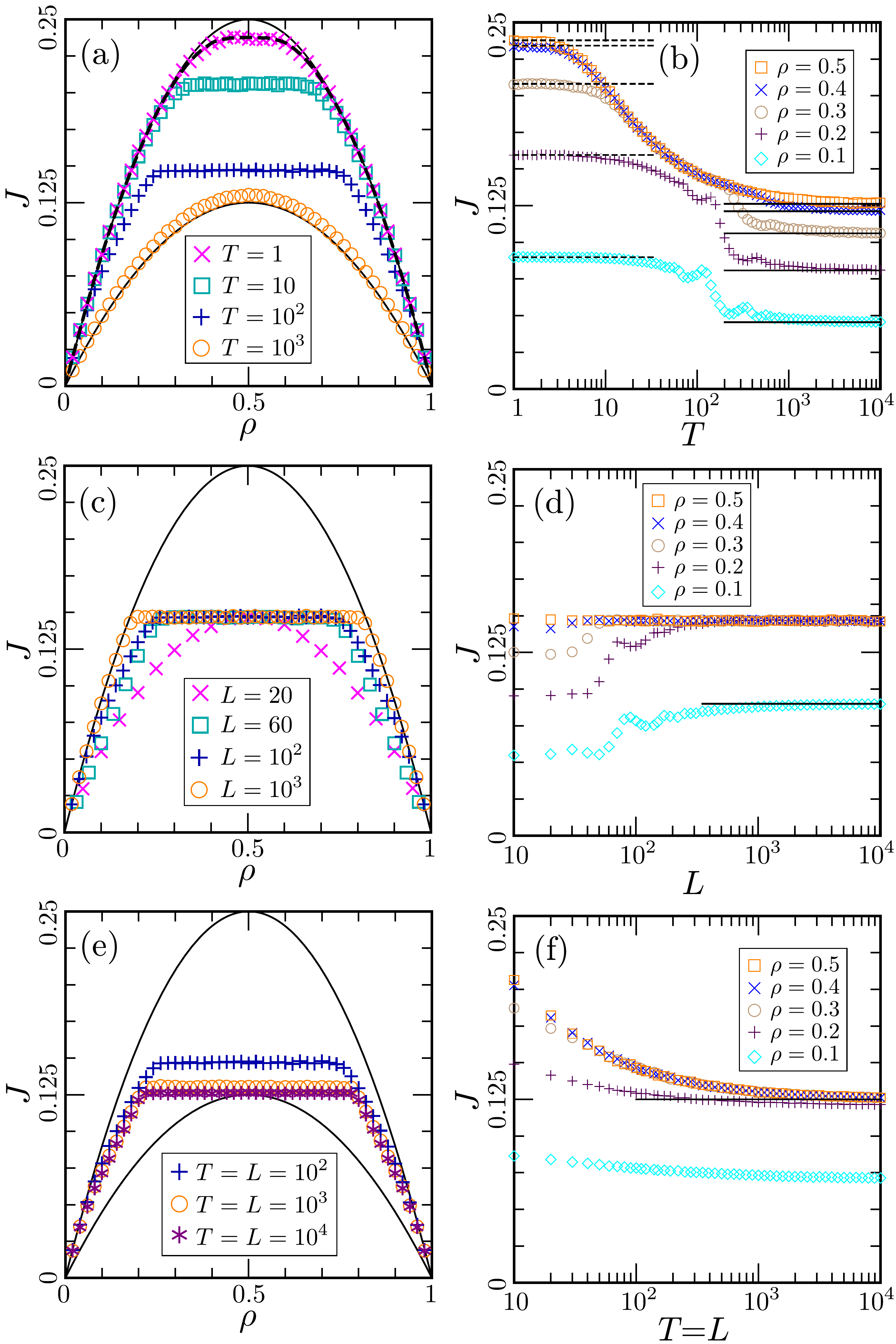}
 \end{center}
 \caption{\label{fig:fundamental-diagrams}
Currents in the cases where
 (a,b) $ L=100 $, 
 (c,d) $ T=100 $, 
 and (e,f) $ T=L $. 
 Simulations were performed on a ring with one traffic light ($ g=0.5 $), 
 and averaged over $10^5 \le t\le2\times 10^5 $. 
The solid lines in (a, c, e) are parabolas $ \rho(1-\rho) $ and $g\rho(1-\rho)$,
those in (b, d, f) correspond to the values in the limits $ T\to \infty $,
$ L \to \infty $, and $ T=L \to \infty $, respectively. 
The dashed line in (a,b) correspond to the current of the JL model with the defect site with rate $r=0.5$.}
\end{figure}

We are interested in the current as well as the density profiles. The fundamental diagram (the relationship between the current $ J $ and the density $ \rho $) depends on the system length $ L $ and the period $T$ of the traffic light, as illustrated in Fig.~\ref{fig:limits}. Simulation results are also summarized in Fig.~\ref{fig:fundamental-diagrams}. In statistical physics, analyses in the ``thermodynamic'' limit $ L\to \infty $ are usually considered to be important. This would be a reference case in our model, but simulations with finite $L$ are also relevant to real traffic. 

Our traffic light model is very similar to the TASEP with a ``blockage'', which was introduced by Janowsky and Lebowitz (JL) \cite{bib:JL:92}. The JL model contains one ``defective'' bond between sites $L$ and 1 where a reduced transition rate $r<1$ is independent of time. Within the mean-field approximation, the fundamental diagram is found to be 
\begin{align}
 \label{eq:J-rho-JL}
 J \approx 
 \begin{cases}
 \rho (1-\rho) & ( \rho \le \rho ^* \vee 1- \rho^* \le \rho) , \\ 
 J^* & ( \rho^* < \rho < 1- \rho^* ) . 
 \end{cases}
\end{align} 
with the critical density $ \rho^* = \frac{r}{1+r } $ 
and the maximal current $J^*= \frac{ r }{ (1+r)^2 } $. 
In particular, for $ \rho^* < \rho < 1- \rho^* $, a shock is localized around the site $S$, 
which is determined by 
\begin{align}\label{eq:determine-S}
 \rho^* S + (1-\rho^*) ( L - S ) = L \rho . 
\end{align}
It separates the density profile into low- and high-density regions ($ 0<j<S $ and $ S<j<L $, respectively). 
The true current and density profile of the JL model seem to qualitatively agree with the above predictions. Because the exact solution is lacking, it is still a challenging problem whether the phase transitions are mathematically true in the limit $ L\to \infty $ \cite{bib:CLST:13}. In the limit $T\to 0$ our model is equivalent to the JL model (Fig.~\ref{fig:limits}), which can be understood as follows. Assume that cars always want to jump from site $ L $ to site 1 with rate 1. But jumps are only allowed when the signal is green. When $T$ is very small, i.e. in the very high frequency of the light, the acceptance of the jump is almost stochastic with probability $ g $. 
 In Fig.~\ref{fig:fundamental-diagrams} (a), we observe that the current $J$ (for $L=100$) already agrees with the JL model for $T=1$. From Fig. \ref{fig:fundamental-diagrams} (b), we expect that, for given $g$, the current takes a supremum in the JL limit $T\to 0$ \cite{bib:Nishimori:13}. We note that this case is not relevant to real traffic. In reality no car can start to move when the frequency of the light is too fast. Therefore slow-to-start effect is needed, when one wants to discuss optimization problem by adjusting the parameter $T$. Furthermore we should take into account a minimum period such that cars can go through the intersection. However the JL model provides qualitatively essential features of our model, e.g. similar fundamental diagrams in some cases and existence of a localized shock in the density profile $ \rho_j $, as defined in Eq.~\eqref{eq:rhoj=}. 
 
In the opposite limit $T\to \infty$ with $ L $ fixed, the current is easy to calculate. In the green phase the system can be regarded as the homogeneous TASEP on a ring (there is sufficient time for relaxation), hence the current $ \simeq J_\text{G} = \rho(1-\rho)\frac{L}{L-1}$. In the red phase all the cars make a queue behind the traffic light and no car can move, hence $J_\text{R} = 0$. For the total current we have
\begin{align}
 \lim_{ T\to \infty } J = gJ_\text{G}+(1-g)J_\text{R}=g\rho(1-\rho) L/ ( L-1 ) ,
\end{align}
 neglecting the transient currents from the red phase to the green phase and from green to red. The factor $L/ ( L-1 ) $ corresponds to the exact finite-size effect, and we have 
\begin{align}
 \label{eq:LinftyTinfty} 
 \lim_{L\to \infty}\lim_{T\to \infty} J = g \rho(1-\rho) . 
\end{align}

\begin{figure} 
 \begin{center}
 \includegraphics[width=76mm]{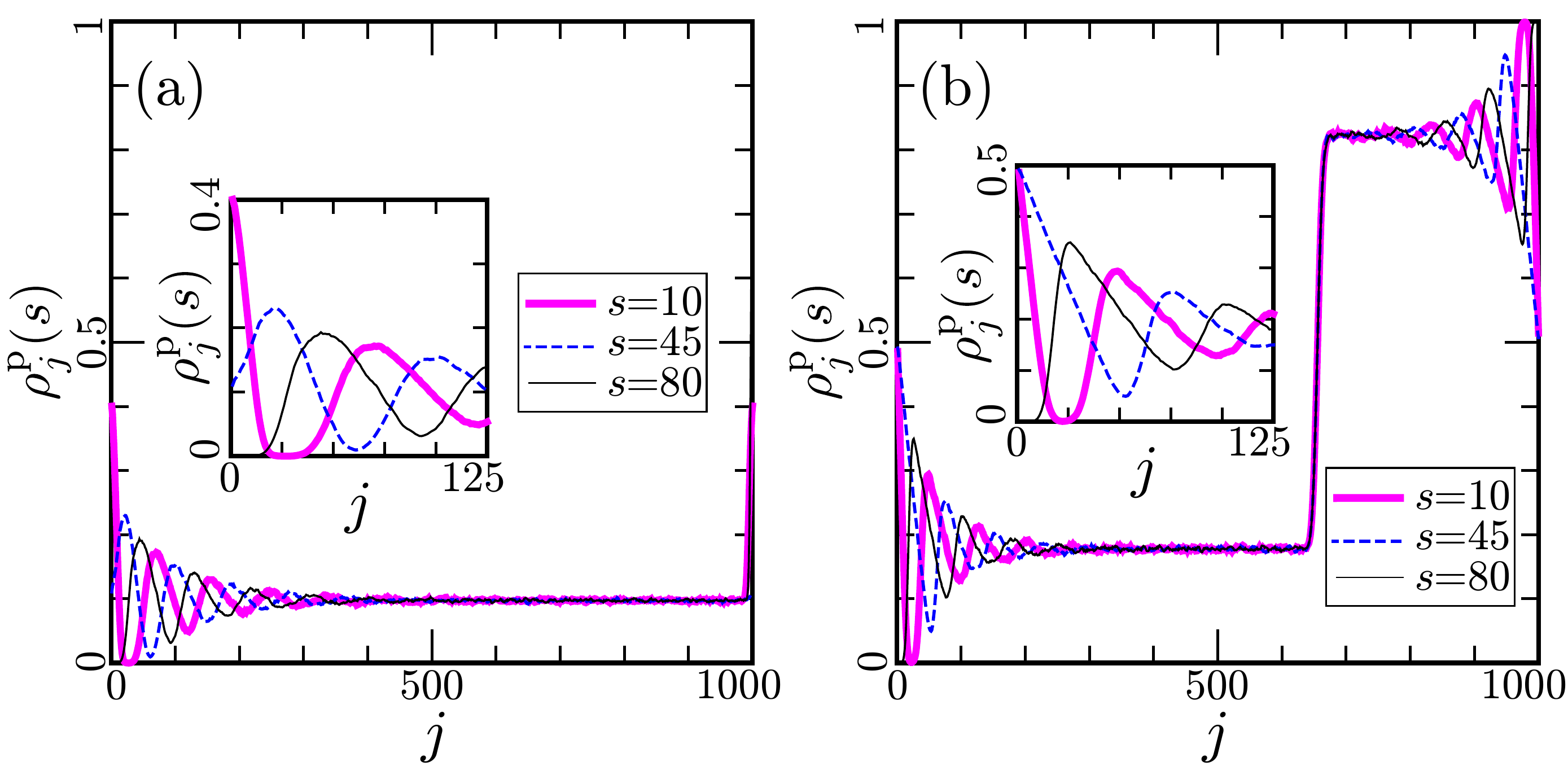}
 \end{center}
\caption{Periodic density profiles $ \rho^{\rm p }_j (s)$ 
 for $T=100$, $L=1000$, and $ g=0.5$. 
 The global density is chosen as (a) $\rho = 0.1$ and (b) $\rho= 0.4$.
 Insets: enlarged view near $j=0$ is shown. 
 }\label{fig:DPT100L1000}
\end{figure}

One can consider another limit, i.e. $L\to \infty$ with $ T $ fixed, 
see Fig.~\ref{fig:fundamental-diagrams} (c,d). The fundamental diagram has conjecturally a similar structure to the JL model, i.e. 
\begin{align}
 \lim_{ L\to \infty } J = 
 \begin{cases}
 \rho (1-\rho) & ( \rho \le \rho ^* \vee 1- \rho^* \le \rho) , \\ 
 J^* & ( \rho^* < \rho < 1- \rho^* ) , 
 \end{cases}
\end{align} 
 where the plateau value $ J^* $ and critical density $ \rho^* $ are different from the JL model. 
In particular we expect 
\begin{align}
 \label{eq:TinftyLinfty}
 & \lim_{T\to \infty} \lim_{L\to \infty} J = 
 \begin{cases}
 \rho (1-\rho) & ( \rho \le \rho^* \vee 1- \rho^* \le \rho) , \\ 
 \frac g 4 & ( \rho^* < \rho < 1- \rho^*) , 
 \end{cases}
\end{align} 
 where $\rho^*$ is the smaller solution to $ \rho^*(1-\rho^*) =\frac g 4 $, i.e. $ \rho^* = \frac{ 1-\sqrt{ 1-g } }{2} $. Note that the order of the two limits here is different from \eqref{eq:LinftyTinfty}. The plateau current $\frac g 4 $ is explained as follows: in the red phase, a long queue $ \dots 111$ is formed behind the light, and there is a large space without cars after the light. Therefore in the green phase, the car current through the traffic light becomes $ \frac 1 4 $. As analyzed in the semi-infinite lattice \cite{bib:PSS:08}, the density profile $ \rho^\text{p}_j (t) $ has a similar sawtooth structure. In the low- (resp. $\text{high-)}$ density case, only one sawtooth profile appears after (resp. before) the light, see Fig. \ref{fig:DPT100L1000} (a). On the other hand, in the intermediate density case, two sawtooth-like profiles appear after and before the light, see Fig.~\ref{fig:DPT100L1000} (b). Far from the light, they converge to $ \rho^* $ and $ 1- \rho^* $, respectively, and the shock position $S$ is given by the same equation \eqref{eq:determine-S} as the JL model.

Let us consider the case where $L$ and $T$ are in the same order, say, 
$ 0.5 < L/T <2 $. The relevant case to real traffic would be $ L \sim 100, T\sim 100 $. 
A remarkable fact is that the current does not always monotonically decrease as $ L $ or $ T $ increases, when the density is low, and $L$ and $T$ are in the same order, see Fig.~\ref{fig:fundamental-diagrams} (b,d). These `fluctuations' indicates that, for given $L$, one can optimize the flow by changing $T$ in this simple geometry. Similar behavior has also been reported in more sophisticated traffic models with traffic lights \cite{bib:BBSS:01,bib:HH:03}. They can be considered as a signature of a periodic function $J(T)$ which has been observed for a traffic model with smaller noise amplitude \cite{bib:BBSS:01}. In Fig. \ref{fig:fundamental-diagrams} (e), there is still a plateau: 
\begin{align}
 J \approx 
 \begin{cases}
 f( \rho ) & ( \rho \le \rho ^* \vee 1- \rho^* \le \rho) , \\ 
 J^* & ( \rho^* < \rho < 1- \rho^* ) 
 \end{cases}
\end{align} 
with some function $f$. 
In the limit $ L\to \infty $ with $ L/T $ fixed,
we expect that the plateau density becomes again $ J^*= \frac g 4 $, 
\begin{align}
\label{eq:TLinfty}
 \lim_{T,L\to \infty\atop T/L \ \text{fixed} } J = 
 \begin{cases}
 f( \rho ) & ( \rho \le \rho ^* \vee 1- \rho^* \le \rho) , \\ 
 \frac g 4 & ( \rho^* < \rho < 1- \rho^* ) . 
 \end{cases}
\end{align} 
The explicit form of the function $f$ is unknown but 
it should satisfy $ g \rho (1-\rho) < f (\rho ) < \rho (1-\rho) $. 
Note that the limits \eqref{eq:LinftyTinfty}, \eqref{eq:TinftyLinfty} and \eqref{eq:TLinfty} are different. 
\begin{figure}
\begin{center}
 \includegraphics[width=76mm]{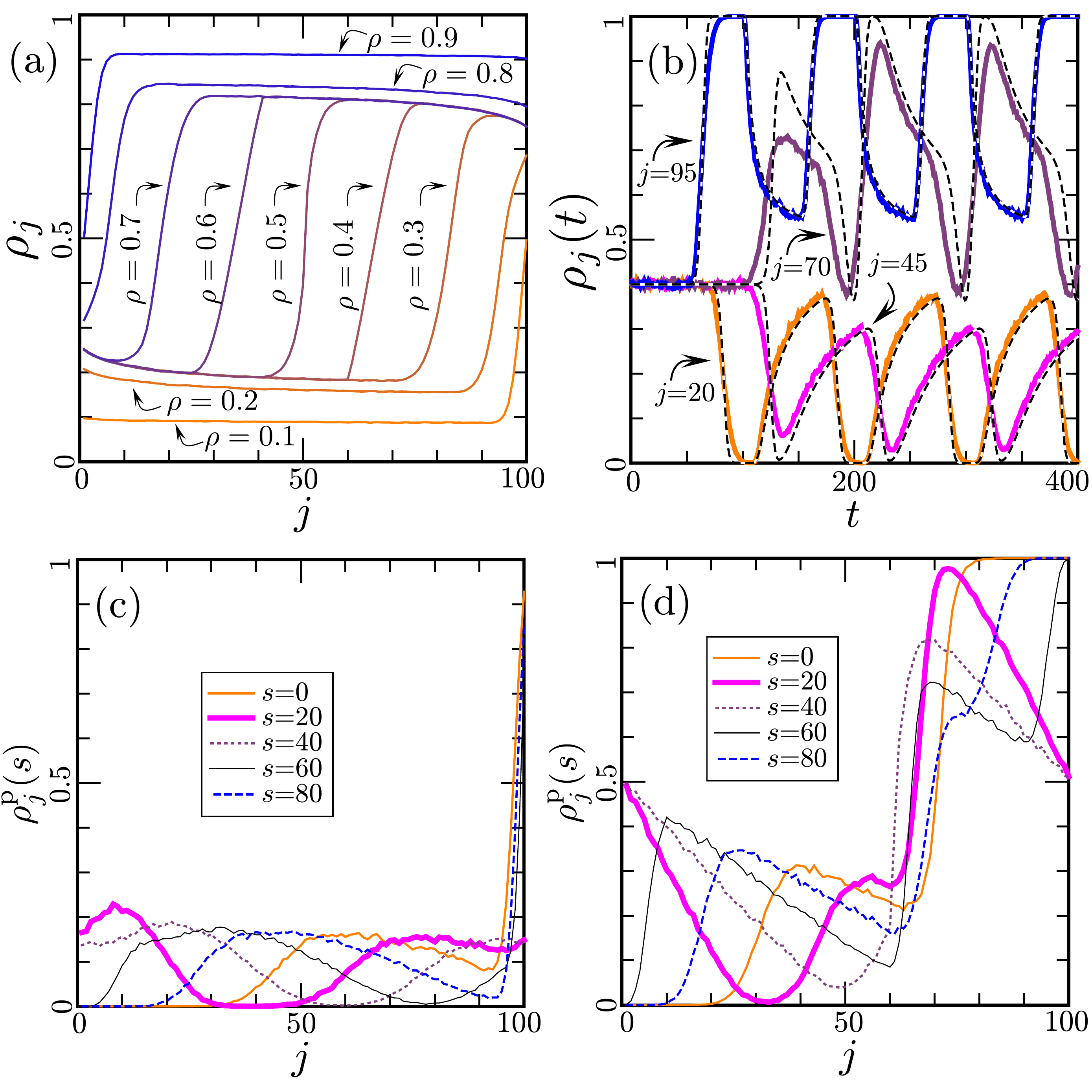}
\end{center}
 \caption{\label{fig:DPs-T=L}
 Density profiles 
 (a) $ \rho_j $ \textit{vs} $j$, 
 (b) $\rho_j(t) $ \textit{vs} $t$ for $\rho = 0.4 $, 
 (c) $\rho^{\rm p}_j(s) $ \textit{vs} $j$ for $\rho = 0.1 $, and 
 (d) $\rho^{\rm p}_j(s) $ \textit{vs} $j$ for $\rho = 0.4 $. 
 The parameters are $ T=L=100$, $g=0.5$. 
 For (a,c,d), we averaged over $ 10^5 \le t \le 10^6 $ of one simulation run. 
 For (b), we averaged over $10^3 $ simulation runs with the initial condition $\rho_j(0)=\rho$. 
 The dashed lines in (b) are numerical results of the mean-field equations \eqref{eq:MF1}--\eqref{eq:MFj}. 
 }
\end{figure}
Let us turn to the density profiles. One observes that the time independent density profile $ \rho_j $, defined in Eq.~\eqref{eq:rhoj=}, exhibits a shock, when the global density
lies in the interval $\rho^*< \rho < 1-\rho^* $, 
 see Fig. \ref{fig:DPs-T=L} (a). In this case only the shock position changes as $\rho $ varies. 
 On the other hand, when $\rho $ is close to 0 or 1, no shock exists. 
Though the fully exact description of the time dependent $\rho_j(t) $ Eq.~\eqref{eq:rhojt=} 
 is difficult, within the mean-field approximation, the rate equations are 
\begin{align}\label{eq:MF1}
 \dot\rho_1(t) &=\sigma(t)\rho_{L}(t) [1-\rho_1(t) ] - \rho_{1}(t) [1-\rho_{2}(t) ] ,\\
 \dot\rho_L(t) &=\rho_{L-1}(t) [1-\rho_L(t) ] -\sigma(t) \rho_{L}(t) [1-\rho_{1}(t) ] , \\
\dot\rho_j(t) &= \rho_{j-1}(t) [1-\rho_j(t) ]- \rho_{j}(t) [1-\rho_{j+1}(t) ]
\label{eq:MFj}
\end{align}
for $2\le j\le L-1 $, where the dots stand for the time derivatives. We numerically probed the time evolution of these equations, by using Euler's method with discretization $\dot\rho_j(t) \to (\rho_j(t + 10^{-4}) - \rho_j(t) ) /10^{-4} $, and the result is compared with a simulation of the original stochastic problem in Fig. \ref{fig:DPs-T=L} (b). The periodicity $\rho_j(t+T) = \rho_j(t) $ can be already observed from time $t\approx 2 T$ as well as qualitative agreement between the simulation and the mean-field approximation \cite{bib:mean-field}. 
In Fig. \ref{fig:DPs-T=L} (c,d), periodic density profiles $ \rho^{\rm p} (s) $ Eq.\eqref{eq:rhopjs} \textit{vs} $j $ are also provided. It is interesting that there is a relation between flat profiles (a) with/without a shock and \textit{fluctuating} profiles (c,d) via $ \rho_j =\int_0^T \rho_j^\text{p} (s) d s $.

\section{Many traffic light problem\label{sec:many}} 
We now consider the general case, i.e. the TASEP on a ring with $n$ traffic lights, where the distance of each pair of successive lights is constant $\ell$ for simplicity. The $i$th light is located at the bond between sites $ i\ell $ and $ i\ell + 1$. In particular the $n$th light is between sites $ L:= n\ell $ and 1. The switch of light $i$ between green $\sigma_i (t) =1$ and red $\sigma_i (t) =0$ phases is defined in terms of the period $ T $, the green phase ratio $0<g<1$ and the offset parameter $ 0\le \Delta_i <1$: 
\begin{align}
\sigma_i (t) =
\begin{cases} 1 &
 ( k+\Delta_i \le \frac t T < k+g+\Delta_i ) , \\
 0 &
 ( k+g+\Delta_i \le \frac t T < k+1+\Delta_i ) 
 \end{cases}
\end{align}
with $k= \lfloor t/T- \Delta_i \rfloor $ showing the cycle number. In other words, the $i$th light periodically becomes green at $ t= ( k +\Delta_i ) T $ and red at $ t= ( k +g +\Delta_i ) T $.  In this work we consider only the case where all the traffic lights have identical values of $ T $ and $g$.

Analogous to the argument for the $n=1$ case, the limit $T \to 0 $ of the model corresponds to the TASEP with $n$ static defective bonds where hopping rates are reduced to $g $. In the opposite limit $T\to \infty$, the current becomes $ J \approx G \rho (1-\rho )$,  where $G$ is the ratio between the period $T$ and the duration when all the lights are green, i.e. $ G = \frac 1 T \int_0^T \sigma_1 (t ) \cdots \sigma_n (t) dt $.
 On the other hand, in the limit $ \ell \to \infty $, the current is expected to be $ \rho (1-\rho) $ for the low and high density cases ($ \rho<\rho^* , \rho>1-\rho^* $), and flat when $ \rho^*< \rho<1-\rho^* $, with some critical density $ \rho^* $. But again we are interested in the case where $ \ell $ and $ T $ are in a same order rather than these limits. 

 The current $J$ depends on the offset parameters $\{ \Delta_i \}$ as well as $T$, $g$ and $ \ell $. In this work we discuss two types of offsetting of traffic lights, i.e. fixed and random offset strategies (see e.g. \cite{bib:Gershenson:05,bib:SIA:13,bib:KSI:16} for other types). In the fixed offset strategy, the difference of the offsets are set as $ \Delta_{i+1} - \Delta_i = \delta \ (\text{modulo 1}) $
for $i=1,2,\dots, n -1 $, with some $0\le \delta < 	1 $. 
The choice of $\delta $ is restricted as
 $\delta = m /n$ ($m = 0,1,\dots, n-1$),
so that no inhomogeneity is caused to the $ n$th light: 
$ \Delta_1 - \Delta_n = \delta $ (modulo 1). 
 The car-hole symmetry $ J |_{\rho\to 1-\rho} = J $ is no longer true
 except for $\delta = 0, 1/2$, but we have an extended symmetry 
$ J |_{\rho\to 1-\rho,\delta \to 1-\delta } = J $. 
 On the other hand, in the random offset strategy, $\Delta_i$ for each $i$ is randomly chosen from the unit interval.

\begin{figure}
 \begin{center}
 \includegraphics[width=76mm]{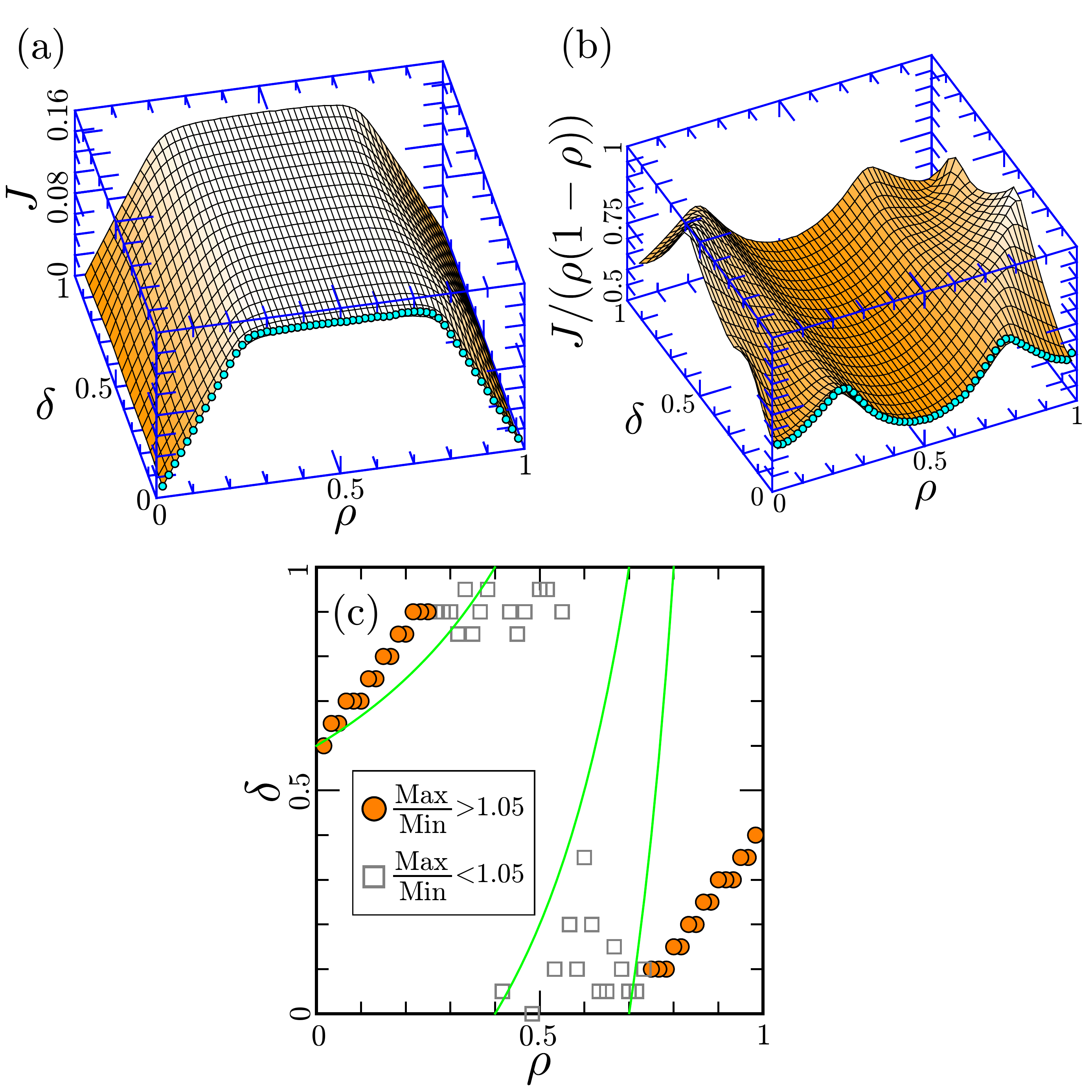}
 \end{center}
\caption{(a) Current $J$ and (b) $ J / [  \rho (1-\rho) ]  $
 \textit{vs} the global density $ \rho $ and the difference of offset parameters $ \delta $ in the fixed offset strategy. The parameters were set as   $ T=100$, $g=0.5$, $ \ell = 60$ and $ n=20$, corresponding to the surface plots. For comparison we plotted data of $ n=1 $ by markers $ \circ $ in the face $ \delta = 0 $. In order to emphasize the effect of $ \delta $ in the low and high density cases, we divide the current by $ \rho (1-\rho) $ in (b). (c) Optimal $ \delta $ \textit{vs} $ \rho $. The curves correspond to the green wave strategy. 
\label{fig:3d-J} }
\end{figure}

Figure~\ref{fig:3d-J} (a) shows the fundamental diagram for $n=20$, $ \ell =60$ and $ T = 100$ in the fixed offset strategy. 
As in the single-light ($n=1$) case, there are regimes $ \rho < \rho^*$ and $1-\rho^*<\rho $,
where the current depends on the global density $ \rho$. When $\rho^*< \rho <1-\rho^*$, the current 
$ J = J^* $ is independent of $\rho $. 
The dependence of this plateau current on $\delta$ is also weak, see Fig.~\ref{fig:3d-J} (a,b). 
On the other hand, in the cases of low and high densities, the dependence on $\delta$ becomes significant.
In Fig.~\ref{fig:3d-J} (c), we plot the value of $ \delta $ which gives maximum of $ J $ for given $ \rho $. 
One may naively think that the so-called green wave strategy, i.e. $\delta = \ell /( v T ) $ (modulo 1) 
with $ v = 1-\rho$, maximizes the current. 
However it should be noted that $v= 1-\rho$ corresponds to the \textit{stationary} velocity of the usual TASEP. Due to the stochasticity of the TASEP, the actual particle velocities and local densities fluctuate 
around their average stationary values. In our case, there is a subtle interplay between density and velocity fluctuations at one hand and the optimal control of traffic lights at the other hand. This is why the optimization of $\delta$ is a non-trivial task and the green wave strategy does not work, as we see in Fig.~\ref{fig:3d-J} (c).

\begin{figure}
 \begin{center}
 \includegraphics[width=76mm]{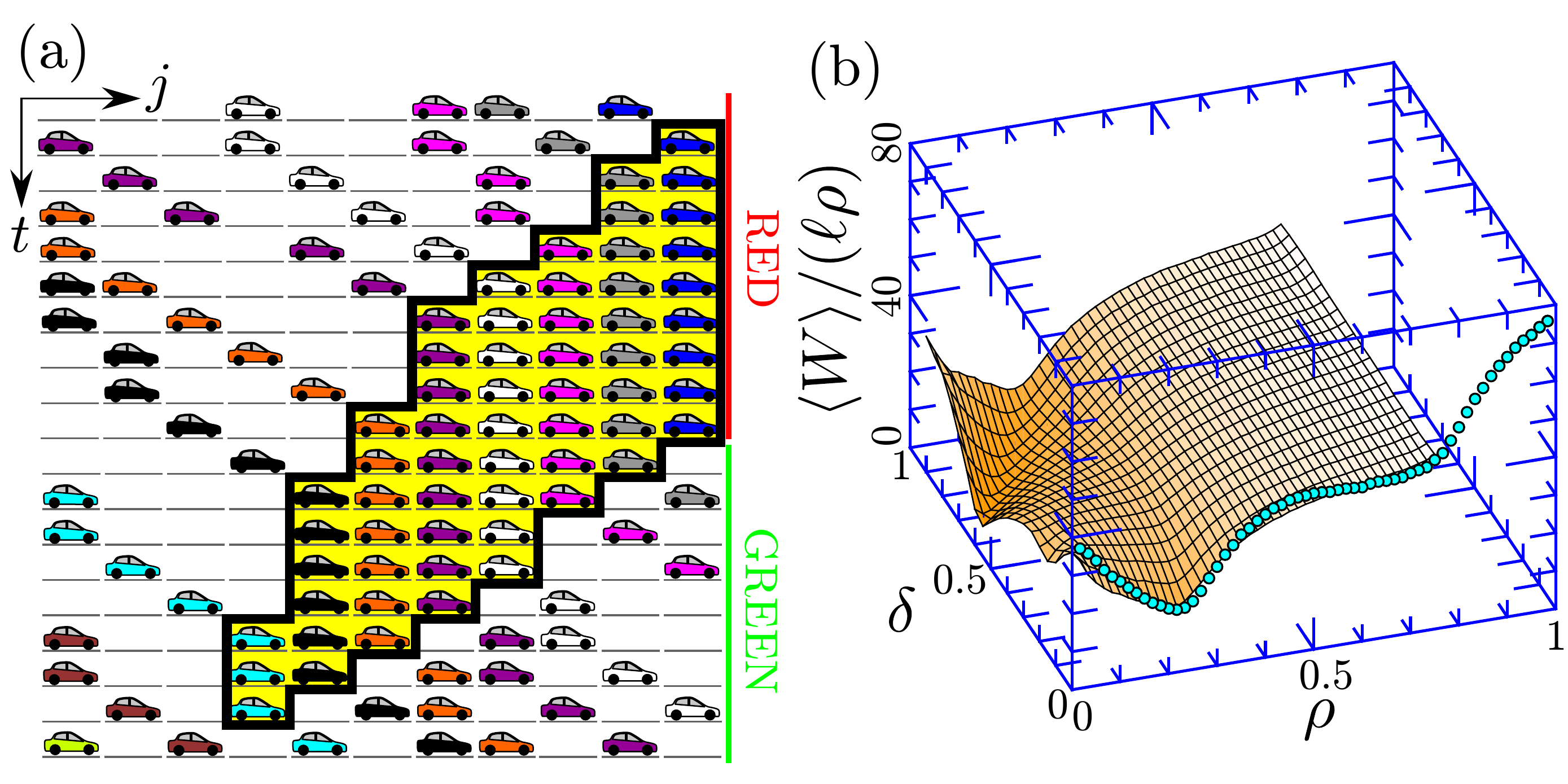} 
 \end{center}
 \caption{(a) Illustration of the total waiting time of each light per period. 
 In this example of a kymograph, the total waiting is the area enclosed by the bold line. 
 (b) Average waiting time $ \langle W \rangle / (\rho \ell) $ \textit{vs} the global density $ \rho $ and the difference of offset parameters $ \delta $. The parameters were set as $T=100$, $g=0.5$, $ \ell =60 $, and $ n=20 $ ($L=1200$), corresponding to the surface plot. For comparison we plotted data of $ n=1 $ by markers $ \circ $ in the face $ \delta = 0 $. 
 \label{fig:W} } 
\end{figure}

Let us turn to investigation on the total waiting time (TWT) \cite{bib:FSS:04,bib:AI:13}. For each car, the waiting time behind a traffic light is defined by the duration from joining a queue to moving again. The TWT $ W_{k,i} $ for the $k$th period of the $i$th light is then the summation of waiting times over all cars in the queue, see Fig. \ref{fig:W} (a). This is also rephrased as the queue length integrated over time. We also denote by $ \langle W \rangle $ the average of TWT over $k$ and $i$. The TWT is one of quantities which we want to minimize in real city traffic. 
In Fig \ref{fig:W} (b) the surface plot corresponds to simulation results of the average TWT normalized by $ \rho \ell $, which is equivalent to average waiting time per car per cycle. When the global density is large ($ \rho > 0.75$), a queue \textit{created} behind the $i$th light in the $k$th red phase can reach the $ (i-1) $th light, and/or it can last still in the $(k+1)$th red phase. In this case we cannot define a TWT per cycle. Therefore we show only meaningful data of $ \rho \le 0.75$ in Fig \ref{fig:W} (b). The dependence on $ \delta $ is again strong in the low density case.

\begin{figure}
 \begin{center}
 \includegraphics[width=76mm]{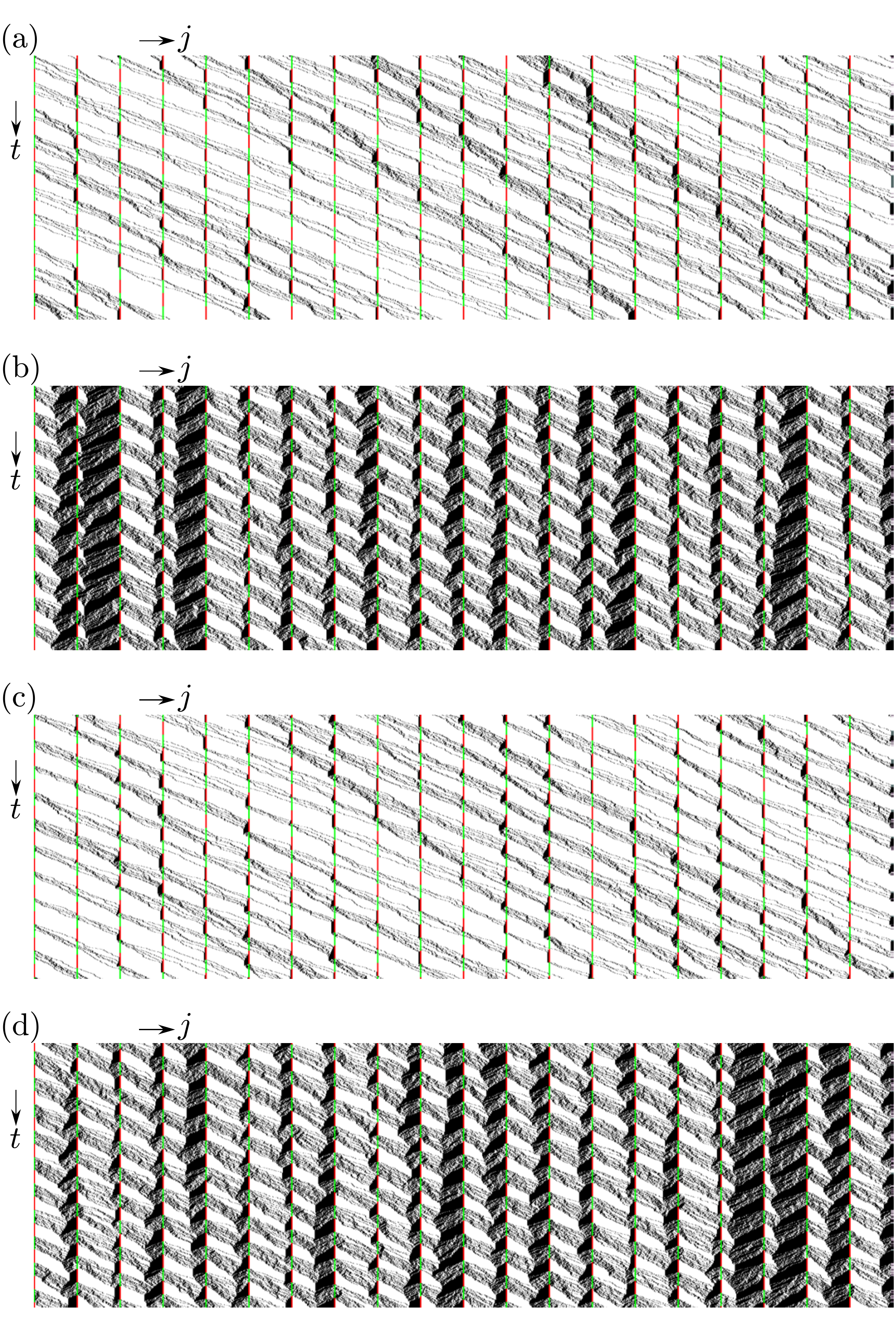} 
 \vspace{-5mm}
 \end{center}
 \caption{Kymographs for the fixed offset strategy with (a) $ (\delta,\rho) = (0.5,0.1) $ and (b) $ (0.5,0.4) $,
 and the random offset strategy with (c) $ \rho=0.1 $ and (d) $ \rho=0.4 $. 
 The other parameters are $ T=100$, $g=0.5$, $ \ell = 60$, and $ n=20$. \label{fig:kymos} }
 \end{figure}

In kymographs, Fig~\ref{fig:kymos}, we observe the following facts. For the low density case (a,c), $ W_{ k,i } $ is fluctuating with respect to $k$, but we partially see that packets are propagated. On the other hand in the plateau current region (the intermediate density regime) (b,d), $ W_{k,i} $ highly depends on $i$ in a time window, say $ t_1 \le t \le t_1 + 1000 $. This property is true for the fixed offset strategy even though there is no inhomogeneity among the lights, i.e. when $ W_{k,i} $ is large (as compared to other lights), $ W_{k+1, i} $ is also large. (This dependence on $i$ disappear when we consider the average over $k$ in a much longer time window.)

\begin{figure}
\begin{center}
 \includegraphics[width=76mm]{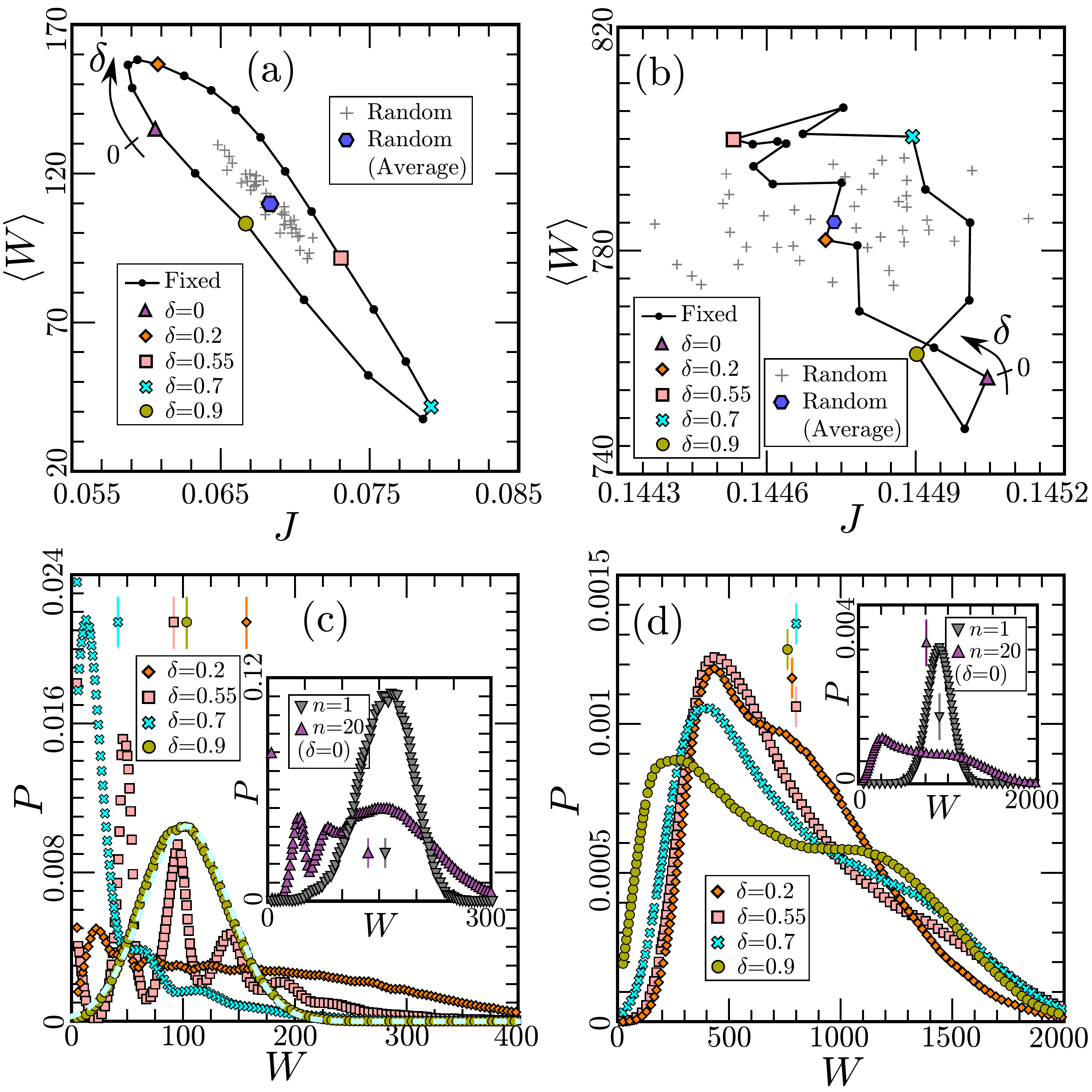}
\end{center}
 \caption{\label{fig:JTWT-TWTdistribution} 
 Plots of $ ( J, \langle W \rangle ) $ as $ \delta$ varies for global density 
 (a) $ \rho = 0.1 $ and (b) $ 0.4 $. 
 We have set the other parameters as $ \ell = 60 $, $ g =0.5 $, $ T=100 $, and $ n=20$. For the fixed offset strategy, we varied the value of $\delta$ as $ 0, 0.05,0.1 \dots, 0.95$. The small dots $ \bullet $ (and partially the markers with various shapes) were obtained by averaging over $10^5 \le t\le 10^6 $. The lines are guides for eyes. For the random offset strategy, each marker $ + $ was obtained by averaging over $10^5 \le t\le 2\times 10^5 $ of one simulation run with a randomly chosen set of parameters $ \{ \Delta_i \} $. Furthermore we averaged 40 simulation runs of the random offset strategy. (c) and (d) show probability distributions of the average TWT for $\rho=0.1$ and 0.4, respectively, in the fixed offset strategy, where the average values are indicated by markers with a vertical bar. In the insets, we compare the distributions in the cases of $ n=1$ and $ n=20$ with $\delta =0$. The dashed line in (c) is the Gaussian fitting for $ ( \rho,\delta) =(0.1, 0.9 ) $.} 
\end{figure}

In the low density regime, there is the following tendency (agreeing with our intuition) as $\delta$ varies: when the TWT is minimized (resp. maximized), the current is maximized (resp. minimized), see Fig. \ref{fig:JTWT-TWTdistribution} (a) for example with $ \rho =0.1$. However the relation between $ J $ and $\langle W \rangle $ is not a perfect one-to-one correspondence. On the other hand, in Fig. \ref{fig:JTWT-TWTdistribution} (b), we provide plots for $ \rho =0.4$ as an example in the intermediate density regime. We cannot observe a clear tendency of the relationship between $ J $ and $ \langle W\rangle $, as their ranges are too small. 
 For comparison, we plot $( J, \langle W \rangle ) $ of the random offset strategy in Fig. \ref{fig:JTWT-TWTdistribution} (a,b). We notice, in (a), that the curve made by the fixed offset strategy encloses 
 the samples of random offset strategy with randomly chosen parameters $ \{ \Delta_i \} $ as well as the average over $ \{ \Delta_i \} $. For intermediate densities $\rho^*< \rho <1-\rho^*$, there is no big difference in $ J $ as $ \delta $ varies, see (b).

Let us investigate the probability distribution $ P(W) $ of the TWT. One may naively expect that the distributions of the cases $ n=1$ and $ n=20 $ with $\delta =0$ are similar to each other. However this guess fails, see the insets of Fig. \ref{fig:JTWT-TWTdistribution} (c,d), where we observe very different curves. For $ \rho = 0.1 $, we observe various types of distributions in Fig.~\ref{fig:JTWT-TWTdistribution} (c). For $ \delta=0.2 $ except for the vicinity of $ W=0 $, the distribution is almost flat, as compared to other values of $\delta $. For $ \delta=0.55 $, we observe a strong oscillation, where peaks appear with a period slightly smaller than $ g T= 50 $. For $ \delta =0.7 $, the distribution is exponential-like (but with a peak at a positive $W $). For $ \delta =0.9 $, a simple Gaussian fitting agrees with the simulation data. On the other hand, for $ \rho =0.4 $ (with $n=20 $), $ P(W) $ does not drastically change its form as we change the value of $ \delta $, see Fig. \ref{fig:JTWT-TWTdistribution} (d).

\section{Summary and conclusions\label{sec:conclusions}} 
In this work we analyzed the TASEP with dynamic defective bonds which correspond to the traffic lights at intersections. Our model can be considered as an approach to the regulation of traffic flow on arterial roads in urban areas. We explored possible optimization strategies of the flow and waiting time behind a traffic light. Our choice of the traffic model, i.e. the continuous-time TASEP, leads to rather strong fluctuations of cars' velocities.
 This implies that the ballistic motion of cars, which might be relevant at low densities, is not considered by our approach. These fluctuations limit the efficiency of the traffic-light optimization schedules based on the typical traveling time between two intersections. In our systematic approach, we started with the single-light problem. We saw that the fundamental diagrams quantitatively depend on the parameters of signalization as well as the system size. Although these time-average fundamental diagrams are qualitatively similar to that of the JL model, the time-periodic rate at the intersection gives interesting phenomena in various types of density profiles. In particular the temporary density profile nicely reflects formation and relaxation of a queue behind the light, which enables us to optimize the flow by tuning the period of the traffic light. In contrast to deterministic discrete-time models, even at low densities it is not possible to avoid queueing of cars completely, but the impact of the signal period and system size is significant. 
Another interesting observation is that the average of the time-periodic profile over one cycle can exhibit a shock.

For the many-light problem, we measured the total waiting time of cars behind traffic lights,
and explored relations to the flow. We found that, for the low density regime, the distribution of the total waiting time takes various forms depending on the offset parameter $ \delta $ of the fixed offset strategy. Moreover the flow and the waiting times are correlated. When the density becomes larger, the flow and the waiting time cannot be controlled by the offset parameters, and the correlation between the flow and the waiting time becomes weaker.

Our results refer to the steady state of a periodic system, which we characterized in some details. We believe that our approach sets a firm ground for the analysis of more sophisticated traffic models as well as in other geometries. 
The step toward more complex lattices has already been made, for example, at an intersection of two perpendicular segments \cite{bib:FNYI:09,bib:FB:11,bib:FIN:13,bib:LJMXW:14,bib:ZJHJ:16} or in more complicated networks \cite{bib:HM:09,bib:SKMY:09,bib:GGR:11,bib:ZGG:13,bib:KMMZG:16}. 
Despite the relevance of these studies for city traffic, in these works traffic optimization has been understood as optimization of the flow. For realistic city traffic, however, it is also important to understand how the cars are distributed in the network as well as to optimize traffic flow with respect to the drivers' waiting times. Although we addressed these issues in a rather simple geometry, we observed that the density of cars is strongly varying at different sections of the roads. This observation is of great relevance for city traffic since a queue on a main road may block a whole section of the city network. 

\vspace{1cm}
\appendix
\section{Proof of the periodic stationarity\label{sec:proof}}
Here we prove that (the ensemble average of) any quantity $Q(t) $ converges to a periodic function 
with period $T$, as $t \to \infty$. 
We consider only the single-light problem, but the proof can be generalized to the many-light problem. 
 The probability distribution $ | P(t) \rangle $ at time $t $ is evolved by 
the master equation in continuous time \cite{bib:SCN:11} 
\begin{align}
 |\dot P(t) \rangle =& M(t)| P(t) \rangle , \quad 
 M(t) = \begin{cases} M_\text{p} & (t' \le g ), \\ M_\text{r} & (t' > g), \end{cases}
\end{align}
where $ t':= t/T- \lfloor t/T \rfloor $, and $M_\text{p}$ and $M_\text{r}$ are transition rate matrices of the TASEPs with the usual periodic and reflecting boundary conditions, respectively.
The series $ \{ | P( \kappa T ) \rangle \}_{ \kappa \in \mathbb Z_{\ge 0} } $ obeys a discrete-time Markov process 
$ | P \textbf{(} (\kappa + 1)T \textbf{)} \rangle = \mathcal M | P( \kappa T ) \rangle$ 
with the transition probability matrix $ \mathcal M = e^{M_\text{r} (1-g) T }e^{M_\text{p} g T } $. 
 This process has a stationary state $ | P_\text{st} \rangle $,
 i.e. $ | P(\kappa T) \rangle = \mathcal M^\kappa | P(0) \rangle \to | P_\text{st} \rangle $ 
in the limit $\kappa \to \infty$,
which satisfies $ | P_\text{st} \rangle = \mathcal M | P_\text{st} \rangle $. 
Using the notation $ s' = s/T - \lfloor  s/T \rfloor  $, we find 
\begin{align}
& | P( \kappa T + s )\rangle \\
& = \begin{cases}
 e^{M_\text{p} s' T } \mathcal M^{ \lfloor  s/T \rfloor + \kappa }| P (0) \rangle & ( s' \le g ) \\
 e^{M_\text{r} (s' -g ) T} e^{M_\text{p} g T } \mathcal M^{ \lfloor  s/T \rfloor + \kappa }| P (0) \rangle & ( s' > g ) 
 \end{cases} \\ 
& \to \begin{cases}
 e^{M_\text{p} s' T} |P_\text{st} \rangle & ( s' \le g ) \\
 e^{M_\text{r} (s' -g)T } e^{M_\text{p} g T } | P_\text{st} \rangle & ( s' > g ) 
 \end{cases} 
 \label{eq:last-equation}
\end{align}
as $ \kappa \to\infty $ ($\kappa\in\mathbb Z_{\ge 0}$). 
We denote Eq.~\eqref{eq:last-equation} by $ | P_\text{st} (s) \rangle $ 
satisfying $ | P_\text{st} (T + s) \rangle= | P_\text{st} (s) \rangle $. 
We find that any quantity $ Q(\kappa T + s) = \langle \mathcal Q | P ( \kappa T + s ) \rangle $
converges to $ \langle \mathcal Q | P_\text{st} ( s ) \rangle $, which is also periodic, $ \langle \mathcal Q | P_\text{st} (T+ s ) \rangle= \langle \mathcal Q | P_\text{st} ( s ) \rangle $.


\begin{thebibliography}{00}
\bibitem{bib:mean-field} 
 In our case, the underlying traffic-model is the TASEP. On a simple periodic lattice the stationary state is given by a product measure. Therefore we expect rather weak particle-particle correlations for our system as well, which leads to
the good agreement between simulations and the mean-field theory.
However, we also observe small quantitative differences in Fig. 5(b);
see e.g. \cite{bib:FKT:08} for a precise analysis of correlations
in a TASEP with some defects. In the case of more sophisticated traffic models, the
agreement between mean-field predictions and simulations is expected to be not as good as in the present case.
\end{thebibliography}
\end{document}